\documentclass[aps,preprint,showpacs,preprintnumbers,amsmath,amssymb]{revtex4}
\usepackage{graphicx}
\usepackage{url}
\begin{document}
\def\beq{\begin{equation}}
\def\eeq{\end{equation}}
\def\beqn{\begin{eqnarray}}
\def\eeqn{\end{eqnarray}}

\def\r {{\bf r}}
\def\x {{\bf x}}
\def\y {{\bf y}}

\def\n {{\bf n}}
\def\d {{\bf d}}

\def\C {{\bf C}}
\def\U {{\bf U}}
\def\D {{\bf D}}
\def\T {{\bf T}}
\def\O {{\bf O}}
\def\i {{\bf i}}
\def\n {{\bf n}}

\def\a {\alpha}
\def\b {\beta}
\def\g {\gamma}
\def\d {\delta}

\def\bk {\bar k}
\def\bq {\bar q}
\def\bs {\bar s}
\def\bl {\bar l}
\def\bj {\bar j}
\def\bM {\bar M}

\def\bcalH {\mbox{\boldmath $\mathcal H$}}
\def\brho {\mbox{\boldmath $\rho$}}
\def\bphi {\mbox{\boldmath $\phi$}}

\title{A novel general mapping for bosonic and fermionic operators in Fock space}
\author{Alexej I.\ Streltsov\footnote{E-mail: Alexej.Streltsov@pci.uni-heidelberg.de}}
\author{Ofir E.\ Alon\footnote{E-mail: Ofir.Alon@pci.uni-heidelberg.de}} 
\author{Lorenz S.\ Cederbaum\footnote{E-mail: Lorenz.Cederbaum@pci.uni-heidelberg.de}} 
\affiliation{Theoretical Chemistry, Institute of Physical Chemistry,
Heidelberg University, Im Neuenheimer Feld 229, 69120 Heidelberg, Germany}
\date{\today}
\begin{abstract}

In this paper we provide a novel and general way to construct the result
of the action of any bosonic or fermionic operator represented in second
quantized form on a state vector, without resorting to the matrix
representation of operators and even to its elements.
The new approach is based on our proposal to compactly enumerate the configurations
(i.e., determinants for fermions, permanents for bosons)
which are the elements of the state vector. This
extremely simplifies the calculation of the action of an operator on a
state vector. The computations of statical properties and of the
evolution dynamics of a system become much more efficient and
applications to systems made of more particles become feasible. Explicit formulations are
given for spin-polarized fermionic systems and spinless bosonic systems,
as well as to general (two-component) fermionic systems, two-component
bosonic systems, and mixtures thereof.

\end{abstract}
\maketitle

\section{Introduction}\label{Sec.I}

The well-known time-dependent Schr\"odinger equation governs the dynamics of many-particle quantum systems
in different fields of modern physics \cite{Book_dynamics1,Book_dynamics2,Nuclear_book,Book_dynamics3,Pit_Stri_book,Book_dynamics4,Book_MCTDH}.
To define a quantum system means to specify its Hamiltonian.
A quantum system is considered to be made of interacting constituent parts,
usually treated as point particles with some known characteristics like mass and charge.
If the interaction potential between these particles is known then the Hamiltonian of the quantum system is defined.
Eigenvectors and eigenvalues of this Hamiltonian provide complete description 
of all the properties of the considered isolated quantum system.

In general, exact solutions for many-particle Hamiltonians are not known and, therefore, different numerical approaches
and techniques are in use \cite{Book_dynamics1,Book_dynamics2,Nuclear_book,Book_dynamics3,Pit_Stri_book,Book_dynamics4,Book_MCTDH,CI}.
The basic and simplest approach is to represent the {\it unknown} many-body wavefunction 
as a linear combination of some {\it known} many-body wavefunctions, i.e., to expand the solution in a known basis set. 
To solve the problem means to find the expansion coefficients or their evolution
depending on whether one performs time-independent or time-dependent studies.

For a quantum system made of indistinguishable particles, e.g., fermions or bosons,
along with the Hamiltonian one has also to specify the quantum statistics of the system.
Statistics enters the many-body solution via the basis set used:
if each and every basis function fulfills a defined quantum statistics then
any linear combination of the basis functions also possesses the same statistics.
From the other end statistics reduces the required size of the Hilbert subspace,
i.e., it allows to operate with a smaller number of many-body basis functions.

In this paper we specifically deal with a system of $N$ identical particles. 
We utilize the commonly used many-body basis functions for bosons, permanents \cite{Pit_Stri_book} 
constructed as a symmetrized product of $N$ one-particle functions or orbitals, and
for fermions, determinants which are antisymmetrized Hartree products of orbitals \cite{CI}.
All many-body basis functions  (determinants or permanents) are attributed to configurations
where $N$ particles reside in $M$ orbitals; for bosons $M\ge1$ for fermions $M\ge N$.
For orthonormal many-body basis sets one can associate the number of the many-body basis functions used 
with the size of the respective Hilbert subspace.
Intuitively it is clear that a larger number of independent many-body basis functions
provides a better description of the many-body solution.
The expansion spanned by all possible permutations for fixed $M$ and $N$ is referred to as a 
full configurational expansion or a complete Fock subspace.
If the many-body wavefunction is represented as a linear combination of permanents 
for bosons and determinants for fermions, then the Hamiltonian of the system
as well as any other operator can also be expressed in second quantized form.

Once the finite many-body basis set is specified one can construct the respective Hamiltonian matrix.
Straightforward implementation of the standard quantum mechanical rules requires 
construction and operations with this Hamiltonian matrix.
For example, typical diagonalization or propagation schemes, like the short iterative Lanczos (SIL) \cite{SIL},
utilize as a standard building block the product of this matrix to a corresponding state vector, 
representing the many-body wavefunction.
In this work we propose an absolutely different ``ideology'' which allows to get the required result 
of the action of the Hamiltonian on a state vector
without construction of the Hamiltonian matrix at all.
The proposed novel theory can be effectively and naturally applied to any general operator 
represented in second quantized form. The construction and operation 
with the corresponding matrix is not needed.
We show the generality and applicability 
of this theory to systems of fermions, bosons as well as to the mixtures thereof.

The proposed theory has already been implemented for bosonic systems
\cite{BJJ_arXiv,to_be_submitted} within the multiconfigurational
time-dependent Hartree for bosons (MCTDHB) \cite{MCTDHB}. Applications to
multi-boson long-time dynamics in double-well \cite{BJJ_arXiv} and
triple-well \cite{to_be_submitted} traps have already been performed
successfully. These studies demonstrate that the ideas proposed in this
work are indeed very beneficial and can be considered as a breakthrough in
computing dynamics.

The structure of the paper is as follows.
For the sake of expositional clarity we first consider in section \ref{secII} a system of spin-polarized fermions, 
i.e., the system where all spins of all fermions are identical, so each and every fermionic orbital can be occupied only by one particle.
In subsection \ref{secII.B} we enumerate and address the fermionic configurations in a simple, compact and practical way.
The novel idea how to get the action of one- and two- body fermionic operators on a state vector as well as of the general Hamiltonian 
is explained in subsection \ref{secII.C}.
Subsection \ref{secII.D} deals with other quantities like expectation values of the Hamiltonian and density matrices.
In section \ref{secIII} we consider a system of bosons
and apply the proposed theory to get the result of the action of a
bosonic Hamiltonian on a state vector, without construction of the corresponding Hamiltonian matrix.
In subsection \ref{secIII.B} we first describe how to map the fermionic onto the bosonic configurations.
This allows us to adopt, after small modification, the fermionic enumeration scheme for bosons.
Next, in subsection \ref{secIII.C}, we see how bosonic one- and two-body operators act on a state vector 
and then give explicitly the desired total action of the bosonic Hamiltonian on it 
as well as related quantities like expectation values of the Hamiltonian and other one- and two-body operators.
In section \ref{secIV} we generalize our ideas and findings to multi-component systems.
In subsection \ref{secIV.A} we first deal with general fermions, i.e.,
the system where fermions with up and down spin projections coexist.
Then we demonstrate the applicability of the theory to binary mixtures of bosons in subsection \ref{secIV.B}
and to systems made of spinless bosons and spin-polarized fermions in subsection \ref{secIV.C}.
Extension of the proposed theory to multi-component systems is given in subsection \ref{secIV.D}.
In section \ref{secV} we discuss the practical implementation of the proposed ideas
and section \ref{secVI} summarizes and concludes.

\section{The case of spin polarized fermions}\label{secII}

\subsection{Hamiltonian of interacting systems and the state vector $\left|\Psi\right>$}\label{secII.A}

Let us define the system first.
We consider a general Hamiltonian in the second quantization form
with one-body and two-body interaction terms:
\beq\label{MB_Ham}
 \hat H =
 \sum_{k,q} h_{kq}  b_k^\dag b_q
 + \frac{1}{2}\sum_{k,s,q,l} W_{ksql} b_k^\dag b_s^\dag b_l b_q,
\eeq
where the matrix elements $h_{kq}$ of the one- and $W_{ksql}$ of the two-body operators are assumed to be known.
Three-body and higher-order interaction terms can also be included in an obvious way.
To complete the definition of the system,
one has to specify the commutation relations for the creation and annihilation operators $b_k^\dag$ and $b_q$.
Here, we operate with the systems of indistinguishable fermions and,
therefore, the usual anticommutation relations are fulfilled:
$b_k b^\dag_q  + b^\dag_q b_k = \delta_{kq}$.

We expand the generic state vector of the many-body system
in a linear combination of $N_{\mathit{conf}}$ {\it known} many-body basis functions $|\vec{n}\rangle$:
\beq\label{MB_Psi}
\left|\Psi\right> =
\sum^{N_{\mathit{conf}}}_{\vec{n}}C_{\vec{n}}\left|\vec{n}\right>.
\eeq
Traditionally, Slater determinants are taken as $|\vec{n}\rangle$ for fermionic systems.
Using fermionic creation operators $b^\dag_k$ each Slater determinant is assembled as
\beq\label{basic_determinants}
 \left|\vec{n}\right> =
\left(b_1^\dag\right)^{n_1}\left(b_2^\dag\right)^{n_2}
\cdots\left(b_M^\dag\right)^{n_M}\left|vac\right>,
\eeq
where $n_i$ can be either  ``0'' or ``1'' for spin-polarized fermions.
$\vec{n}=(n_1,n_2,n_3,\cdots,n_M)$ represents the occupations of the orbitals
that preserve the total number of particles $n_1+n_2+n_3+\cdots+n_M=N$,
$M$ is a number of the one-particle functions, here $M \ge N$, 
and $\left|vac\right>$ is the vacuum.

In the above written expansion Eq.~(\ref{MB_Psi}) we did not specify explicitly
the size $N_{\mathit{conf}}$ of the problem, i.e., the size or length of the $\left|\Psi\right>$ vector.
Let us now consider the configurational space spanned by {\it all}
possible distributions of $N$ fermions over $M$ fermionic orbitals, i.e.,
a complete Fock subspace of the respective configurational space.
For spin-polarized fermions the size of such a complete Fock subspace \cite{CI,CC_paper} is
\beq\label{Nconf.f}
N_{\mathit{conf}}=\binom{M}{N},
\eeq
where $\binom{n}{k}=\frac{n!}{k!(n-k)!}$.
On the other hand, $N_{\mathit{conf}}$ is the dimension, i.e., the number of the elements
of any state vector $\left|\Psi\right>$ of the system.

\subsection{Enumeration of the Slater determinants} \label{secII.B}

Now our goal is to provide simple and compact scheme for enumeration of the configurations.
More strictly, we can reformulate this enumeration as a requirement to map $M$ integers $ n_1,n_2,n_3,\cdots,n_M$
characterizing each configuration, i.e., Slater determinant [see Eq.~(\ref{basic_determinants})]
to {\it one} integer addressing it as a coordinate (index) of the state vector.
Clearly, there are many ways to solve this problem.
Here we report on one that utilizes the so-called Combinadic numbers \cite{Combinadic}.

Every fermionic configuration can be represented as a
vector with $M$ components, filled by ``1'' or ``0''.
The number of orbitals $M$ must be larger
than the  number of spin polarized fermions $N$.
Each appearing ``1'' means that the corresponding orbital
is occupied by one fermion and ``0'' means that it is not occupied, i.e., it has a hole.
Since the total number of ``1'' and ``0'' characterizing a configurational
vector, i.e., its length, is $M$, there are $N$ particles -- ``1'' -- and $M_v=M-N$ holes -- ``0''.
Here $M_v$ specifies the number of unoccupied, i.e., virtual orbitals.
The general fermionic configuration reads:
\begin{equation} \label{i-labeling}
 |\underbrace{\overbrace{\overbrace{11111110}^{i_1}1111110}^{i_2}\cdots111110}_{i_{M_v}}1111111\rangle.
\end{equation}
For example, for the system of $N=7$
polarized fermions distributed over $M=10$ fermionic orbitals there are $N=7$ occupied orbitals
and $M_v=M-N=3$ unoccupied ones.
Then, for instance, the fermionic configuration $|1011101011\rangle$ means that
the second, sixth and eighth fermionic orbitals are vacant, i.e., are not occupied,
while the first, third, fourth, fifth, seventh, ninth and tenth orbitals are occupied by one fermion.
The problem of enumeration of $M$ dimensional vectors filled by ``1'' and ``0''
has been successfully solved in the context of Ref.~\cite{Combinadic}; here we utilize it.

For a given system with $N$ fermions and $M$ fermionic orbitals the {\it address} of
every fermionic configuration can be uniquely defined either by
specifying the positions of all the ``1''s, i.e., particles, or
alternatively by giving the positions of all holes, i.e., the ``0''s.
For the sake of definitiveness and without loss of generality
we use the positions of the holes to specify the configurations.
The general configuration defined in
Eq.~(\ref{i-labeling}) is described by $M_v=M-N$ holes placed at positions $i_1,i_2,\cdots,i_{M_v}$.
It is convenient to order the holes: $i_1<i_2<i_3<\cdots<i_{M_v}$. 
The address of this configuration is computed as follows:
\beqn\label{i-numbering}
J(i_1,i_2,\cdots,i_{M_v})=1+\sum_{k=1}^{M_v}\binom{N+M_v-i_k}{M_v+1-k}.
\eeqn
So, each and every fermionic configuration in a fermionic state vector $\left|\Psi\right>$
has its own unique address (index),
defined by the numbers $i_1,i_2,\cdots,i_{M_v}$.
For the above considered fermionic configuration $|1011101011\rangle$
with $N=7$ particles and $M_v=3$ holes, the ``0''s are located at $i_1=2,i_2=6,i_3=8$
when we count their positions from the left [see Eq.~(\ref{i-labeling})].
Using Eq.~(\ref{i-numbering}) we obtain
the address of this configuration $J(2,6,8)=\binom{10-2}{3}+\binom{10-6}{2}+\binom{10-8}{1}+1=65$.
We note that the enumeration scheme provided in Ref.~\cite{Combinadic} uses positions
of the ``1''s counted from the right,
while in our scheme we count the positions of the ``0''s, i.e., holes from the left.
The inverse problem, i.e.,
a restoration of the ``holes positions'' $i_1,i_2,\cdots,i_{M_v}$
according to a given address $J$ also can be solved \cite{Combinadic} if needed.
Our goal is fulfilled -- spin-polarized fermionic configurations are enumerated in an easy and compact way.

Let us summarize,
$N$ spin-polarized fermions distributed over $M$ fermionic orbitals
span the complete subspace of the Fock space of $N_{\mathit{conf}}=\binom{N+M_v}{N}$ configurations, here
$M_v=M-N$ is the number of unoccupied fermionic orbitals.
The dimension of any state vector $\left|\Psi\right>$ of this system is $N_{\mathit{conf}}$. 
Every fermionic configuration Eq.~(\ref{i-labeling}) in the respective Fock subspace is characterized 
by the positions of $M_v$ holes placed at $i_1,i_2,\cdots,i_{M_v}$.
We can attribute a unique address $J$ to each fermionic configuration according to the
rule Eq.~(\ref{i-numbering}): $J=J(i_1,i_2,\cdots,i_{M_v})$.
This Combinadic-based mapping is a one-to-one mapping.
In particular, the index $J$ takes all values between 1 and $N_{\mathit{conf}}$.

Now a general $\left|\Psi\right>$, Eq.~(\ref{MB_Psi}), can be rewritten in a specific form:
\beq\label{J_Psi.f}
\left|\Psi\right> =
\sum^{N_{\mathit{conf}}}_{J=1}C_{J}\left|J(i_1,i_2,\cdots,i_{M_v})\right>=\sum^{N_{\mathit{conf}}}_{J=1}C_{J}\left|J(\i)\right>,
\eeq
where it is explicitly stated that every fermionic configuration, i.e., Slater determinant $\left|J(\i)\right>$ 
is specified by $M_v$ holes placed at $(i_1,i_2,\cdots,i_{M_v})\equiv \i$. It has a unique address
$J=J(i_1,i_2,\cdots,i_{M_v})$ which can be computed according to Eq.~(\ref{i-numbering}).
The index (address) $J$ runs over all $N_{\mathit{conf}}$ configurations.

\subsection{Applying operators to $\left|\Psi\right>$}\label{secII.C}

The Hamiltonian (\ref{MB_Ham}) is defined as a sum of terms,
each of them is a product of creation and annihilation operators --
a pair $b^\dag_kb_q$ for the one-body and a quartet $b^\dag_kb^\dag_sb_lb_q$ 
for the two-body terms, scaled by respective prefactors, i.e., integrals $h_{kq}$ and $W_{ksql}$. 
In principle, any other operator can be represented in a similar way as 
a sum of contribution of one-, two-, or higher-oder terms.
The idea is quite simple:
if one would know the result of action of each of these terms on a state vector $\left|\Psi\right>$,
the sum of all of them would give the required total result of the action of the Hamiltonian on a state vector.
Here we would like to recall that the total number of terms in the Hamiltonian
is defined by the number of the orbitals used.
In the most general case of $M$ orbitals the total number
of the one-body terms is $M^2$, and the total number of the two-body terms is $M^4$.
Any symmetry in the problem, including hermicity, reduces these numbers.

\subsubsection{The action of one- and two-body operators}\label{secII.C.1}

Let us consider the action of a pair $b^\dag_kb_q$ of creation and annihilation operators on every fermionic configuration 
Eq.~(\ref{i-labeling}) of a general state vector $\left|\Psi\right>$.
As a first step we consider a specific, say $b^\dag_2b_3$ term,
that kills a particle in the third orbital and creates a particle in the second one.
For spin-polarized fermions due to Fermi-Dirac statistics
the term $b^\dag_2b_3$ provides a non-zero action only on the subset of configurations
$|n_1,0,1,\cdots,n_M\rangle$ having $n_2\equiv0$, $n_3\equiv1$:
$$
b^\dag_2b_3|n_1,0,1,\cdots,n_M\rangle=(-1)^0|n_1,1,0,\cdots,n_M\rangle,
$$
where $n_i$ can be ``0'' or ``1''.
Similarly for a general $b^\dag_kb_q$ case, $n_k\equiv0$, $n_q\equiv1$:
$$
b^\dag_kb_q|n_1,n_2,n_3,\cdots,n_M\rangle=(-1)^{\sum_{i \in (k,q)} n_i }|n_1,n_2,\cdots,n_q-1,\cdots,n_k+1,\cdots,n_M\rangle,
$$
where $(-1)^{\sum_{i \in (k,q)} n_i }$ is a prefactor
ensuring correct fermionic statistics of the antisymmetrized wavefunction
and the summation $i \in (k,q)$ runs over all occupations $n_i$ between the $k$-th and $q$-th orbitals.
We can interpret this well known result as follows:
operation of any even combination of creation and annihilation operators
on a configuration (determinant) results in the re-addressing of this configuration (determinant) to another one.
Since the occupation numbers of the incoming $|n_1,n_2,n_3,\cdots,n_M\rangle$ and resulting $|n_1,n_2,\cdots,n_q-1,\cdots,n_k+1,\cdots,n_M\rangle$ 
configurations are explicitly known, the numbers of the orbitals with zero occupations, i.e., the holes positions are also available 
and therefore, according to Eq.~(\ref{i-numbering}), we can also compute their addresses in a state vector Eq.~(\ref{J_Psi.f}).
Applying the enumeration scheme (mapping) introduced above one gets:
$$
b^\dag_kb_q \left|J(\i)\right>=(-1)^{d^{kq}_J}\left|J(\i')\right>,\, k \in \i, q\notin \i.
$$
In other words, the $J$-th configuration is re-addressed to a new configuration with index $J'\equiv J(\i')$ with some sign prefactor.
For a given configuration $\left|J(\i)\right>$ and fixed $k,q$ the integer $d^{kq}_J$ is
equal to the number of fermions, i.e., ``1''s located between the $k$-th and $q$-th orbitals: $d^{kq}_J=\sum_{i \in (k,q)} n_i$.
Note that there is no difference in whether to count the number of fermions between the $k$-th and $q$-th orbitals
in the original configuration $J$ or in the resulting $J'$ configuration, namely, $d^{kq}_J=d^{kq}_{J'}$.
It is important to stress that the operation with even combinations of creation and annihilation
operators results in a single-valued re-addressing, i.e., an initial configuration having address $J$ 
is re-addressed to a single configuration with address $J'$. 

Having at hand the result of the action of the pair $b^\dag_kb_q$ on a general configuration (determinant) we
can find out their action on the total state vector [see Eq.~(\ref{J_Psi.f})]:
\beqn\label{one-body_oper.f}
\left|\Psi^{kq}\right>&\equiv& b^\dag_kb_q\left|\Psi\right>=\sum^{N_{\mathit{conf}}}_{J=1}C_{J}b^\dag_kb_q\left|J\right>=
\sum^{N_{\mathit{conf}}}_{J=1}C^{kq}_{J}\left|J(\i)\right>, \\
C^{kq}_{J}&=&\left \{
\begin{matrix}
C_{J^{kq}}(-1)^{d^{kq}_J}; k \not\in \i, q \in \i\\
0; \,\, {\mathit{otherwise}}  \\
\end{matrix}
\right.,\nonumber
\eeqn
where at the last step we have changed variables of the summation index from $J'$ to $J$.
Note, that this closed-form result of the action of the basic one-body operator on a state vector
has been obtained without referring to the matrix representation of the respective operator.
How to understand this result?
Every element $C^{kq}_{J}$ of the resulting vector having address $J$ is
obtained as a product of the $C_{J^{kq}}$ element of the incoming vector
having address $J^{kq}$ scaled by the $(-1)^{d^{kq}_J}$ fermionic prefactor.
The configuration $J^{kq}$ is related to $J$ by making a hole at $k$-th and filling a hole at $q$-th orbitals.
The $b^\dag_kb_q$ term acts only if the index $k$ coincides with one of the holes' positions $(i_1,\cdots,k,\cdots,i_{M_v})$
and the index $q$ with neither of them. If these two conditions are not fulfilled the respective
configuration does not contribute at all to the resulting state vector.
The addresses $J^{kq}=J(i_1,\cdots,k,\cdots,i_{M_v})$ and $J=J(i_1,\cdots,q,\cdots,i_{M_v})$ are computed by Eq.~(\ref{i-numbering}).
In Eq.~(\ref{one-body_oper.f}) the summation runs over the index $J$ implying that for a given $J$ one has to
restore the holes' positions $i_1,i_2,\cdots,i_{M_v}$ first. However, this complication can be easily avoided if,
instead, one starts $M_v$ nested loops running over the positions of the holes $i_1<i_2< \cdots < i_{M_v}$ directly.
In the latter case all $i_1,\cdots,i_{M_v}$ are explicitly available,
as well as the resulting address $J=J(i_1,\cdots,q,\cdots,i_{M_v})$ via Eq.~(\ref{i-numbering}).
Hence, at each step to get the desired element $C^{kq}_{J}$ of the resulting vector
one has to compute one integer $d^{kq}_J$ and apply Eq.~(\ref{i-numbering}) only once
to get the index $J^{kq}=J(i_1,\cdots,k,\cdots,i_{M_v})$ of the respective incoming configuration and its value $C_{J^{kq}}$.

The action of a general ($k \ne s \ne l \ne q$) two-body $b^\dag_kb^\dag_sb_lb_q$ term on 
the incoming state vector $\left|\Psi\right>$ can be obtained using a similar strategy:
\beqn\label{two-body_oper.f}
\left|\Psi^{kslq}\right> &\equiv& b^\dag_kb^\dag_sb_lb_q\left|\Psi\right>=b^\dag_sb_l\left|\Psi^{kq}\right>=
\sum^{N_{\mathit{conf}}}_{J=1}C^{kslq}_{J}\left|J(\i)\right>,\\
C^{kslq}_{J}&=&
\left \{ \begin{matrix}
C_{J^{kslq}}\left(-1\right)^{d^{kq}_{J^{sl}}} \left(-1\right)^{d^{sl}_J}; \,k,s \not\in \i, l,q\in \i\\
0; \,\, {\mathit{otherwise}} \\
\end{matrix} \right.. \nonumber
\eeqn
To get the $J$-th element $C^{kslq}_{J}$ of the resulting state vector one
has to take the element $C_{J^{kslq}}$ of the incoming vector
having address $J^{kslq}$ and multiply it by $\left(-1\right)^{d^{kq}_{J^{sl}}+d^{sl}_J}$ prefactor.
Using Eq.~(\ref{i-numbering}) we compute the addresses of
the resulting $J =J(i_1,\cdots,l,\cdots,q,\cdots,i_{M_v})$
and incoming $J^{kslq}=J(i_1,\cdots,k,\cdots,s,\cdots,i_{M_v})$ configurations.
The configuration $J^{ksql}$ is related to $J$ by making holes at $k$-th and $s$-th and filling holes at $q$-th and $l$-th orbitals.
The $b^\dag_kb_q$ term acts only if the index $k$ coincides with one of the holes' positions $(i_1,\cdots,k,\cdots,i_{M_v})$
and the index $q$ with neither of them.
For a given quartet $k,s,l,q$ every component $J$ of the incoming state vector is characterized
by the integer $d^{kq}_{J^{sl}}+d^{sl}_J$ computed
as a sum of the number $d^{sl}_J$ of fermions located between the $s$-th and $l$-th orbitals of the configuration $J$
and $d^{kq}_{J^{sl}}$ -- between the $k$-th and $q$-th ones of the configuration $J^{sl}$.
The $b^\dag_kb^\dag_sb_lb_q$ term provides non-zero action only on the smaller subset of the configurations having
$n_k\equiv n_s\equiv0$, $n_l\equiv n_q\equiv1$. Therefore, in practical computations one needs to address
only this subset.

\subsubsection{The action of the Hamiltonian}\label{secII.C.2}

Considering configurations as coordinates of the state vector $\left|\Psi\right>$,
we have seen that the action of each term of the Hamiltonian on the state vector re-addresses the
coordinates of the original state vector,
multiplying them by some known prefactors. 
So, instead of constructing the full Hamiltonian matrix and
performing matrix-to-vector multiplications,
we obtained the same result by reordering the components of the incoming state vector according to the
action of every $b^\dag_kb_q$ and $b^\dag_kb^\dag_sb_lb_q$ term
and by multiplying them by the corresponding integrals $h_{kq}$ and $W_{ksql}$ and summing the results up.

The general Hamiltonian Eq.~(\ref{MB_Ham}) is a sum of the one-body $\hat h =\sum_{k,q} h_{kq} b^\dag_kb_q$ 
and two-body $\hat W =\frac{1}{2}\sum_{k,s,q,l} W_{ksql} b^\dag_kb^\dag_sb_lb_q$ terms.
Using the results of the previous subsection we find the total action of all one-body terms 
on the initial state vector $\left|\Psi\right>$:
\beqn\label{H-psi_one-body.f}
\hat h \left|\Psi\right> &=&\sum_{k,q} h_{kq} \left [  b^\dag_kb_q \left|\Psi\right> \right ]
= \sum_{k,q} h_{kq}  \left|\Psi^{kq}\right>=\sum^{N_{\mathit{conf}}}_{J=1}C^{\hat h}_{J}\left|J(\i)\right>, \\
C^{\hat h}_J&=& \sum_{k,q}h_{kq} C^{kq}_J. \nonumber 
\eeqn
Here, all the components $C^{kq}_J$ have been computed above using Eq.~(\ref{one-body_oper.f}).
Analogously, we sum up all the contributions from all two-body terms:
\beqn\label{H-psi_two-body.f}
\hat W \left|\Psi\right> &=&\frac{1}{2}\sum_{k,s,q,l} W_{ksql} \left [b^\dag_k  b^\dag_s b_l b_q\left|\Psi\right> \right ]
=\frac{1}{2}\sum_{k,s,q,l} W_{ksql}  \left|\Psi^{kslq}\right>=\sum^{N_{\mathit{conf}}}_{J=1}C^{\hat W}_{J}\left|J(\i)\right>,\\
C^{\hat W}_{J}&=&\frac{1}{2}\sum_{k,s,q,l} W_{ksql} C^{kslq}_{J}.\nonumber
\eeqn

Finally, the desired action of the Hamiltonian on the state vector $|\Psi\rangle$ is a sum of the two obtained above
vectors in Eqs.~(\ref{H-psi_one-body.f},\ref{H-psi_two-body.f}):
\beqn\label{Hpsi.f}
\hat H | \Psi \rangle&=&\hat h \left|\Psi\right>+\hat W \left|\Psi\right>=
\sum^{N_{\mathit{conf}}}_{J=1} C^{\hat H}_{J} \left|J(\i)\right>, \\
C^{\hat H}_{J}&=&C^{\hat h}_{J} + C^{\hat W}_{J}, \nonumber
\eeqn
which fulfills our initial goal.
This closed-form result of the action of the Hamiltonian on a state vector
is constructed without building the respective Hamiltonian matrix or even without referring to its matrix elements.

\subsection{Other quantities of interest}\label{secII.D}
In the preceding subsections we have successfully derived a simple and straightforward technique to operate with state vectors
without representing the respective operators in matrix form.
In particular, we have utilized the basic fact that the result of action of any operator
represented as a sum of products of creation and annihilation operators
on a state vector is equal to the sum of action of each of these terms.
So, the action of any operator on a state vector (called incoming state vector) results in a new state vector
(called resulting state vector).
Let us now show that the expectation value of the respective operator
can be immediately computed as a dot-product of the incoming and resulting state vectors.

Indeed, according to the standard definition \cite{Loedwin,Yukalov,Corelation},
for a given state vector $\left|\Psi\right>$ the elements of the reduced one-body density matrix read:
\beq\label{rho_ij.f}
\rho_{kq}= \left<\Psi\right| b^\dag_kb_q\left|\Psi\right>\equiv\left<\Psi\right| \left[ b^\dag_kb_q\left|\Psi\right> \right]
=\left<\Psi|\Psi^{kq}\right>=\sum^{N_{\mathit{conf}}}_{J=1}C^*_{J}C^{kq}_{J},
\eeq
where we substitute the result for $b^\dag_kb_q\left|\Psi\right> $ from Eq.~(\ref{one-body_oper.f}).
Thus, the elements of the reduced one-body density matrix $\rho_{kq}$ are obtained as a dot-product
of the incoming $\left|\Psi\right>$ and resulting $\left|\Psi^{kq}\right>$ state vectors.

Similarly, the elements of the reduced two-body density matrix are obtained as a dot-product of the incoming $\left|\Psi\right>$ and
resulting [see Eq.~(\ref{two-body_oper.f})] $\left|\Psi^{kslq}\right>$ state vectors:
\beq\label{rho_ijkl.f}
\rho_{kslq}= \left<\Psi\right| b^\dag_kb^\dag_sb_lb_q \left|\Psi\right>\equiv\left<\Psi\right| \left[ b^\dag_kb^\dag_sb_lb_q\left|\Psi\right> \right]
=\left<\Psi|\Psi^{kslq}\right>=\sum^{N_{\mathit{conf}}}_{J=1}C^*_{J}C^{kslq}_{J}.
\eeq
So, in the discussed scheme the elements of the reduced one- and two-body density matrices appear very naturally.
Moreover, the elements of the reduced three- and higher- order density matrices can be obtained in a very similar way.

Finally, taking the result of the action of the Hamiltonian on an initial state vector
$\hat H | \Psi \rangle$ from Eq.~(\ref{Hpsi.f}), we compute
the respective expectation value of the Hamiltonian as a dot-product as well:
\beqn\label{psiHpsi}
\left<\Psi\right| \hat H \left|\Psi\right>\equiv
\left<\Psi\right| \left[ \hat H \left|\Psi\right> \right]
=\sum^{N_{\mathit{conf}}}_{J=1}C^*_{J}C^{\hat H}_{J}.
\eeqn
We have prescribe here a few expectation values of particular interest in a many-body theory.
Other quantities can be represented and computed in a similar way.

\section{The case of stuctureless bosons}\label{secIII}
\subsection{General remarks}\label{secIII.A}
Here, we deal with a system of $N$ identical interacting bosons.
The terms structureless or spinless are often used 
to specify that the bosons do not have an internal structure.
Our goal is to show that our ideas how to operate with state vectors
without resorting to the matrix representation of the respective operators proposed above for fermions
can naturally be extended and applied to bosonic systems.
The generic Hamiltonian Eq.~(\ref{MB_Ham}) introduced above for the system of spin-polarized fermions
is also applicable to the system of structureless bosons with
the only modification concerning the commutation relations of creation and annihilation operators.
Here, we operate with the systems of indistinguishable bosons and,
therefore, we use the usual commutation relations:
$b_k b^\dag_q  - b^\dag_q b_k = \delta_{kq}$.

The generic bosonic state vector is expanded as a linear combination of $N_{\mathit{conf}}$ known many-body basis functions $|\vec{n}\rangle$:
\beq
\left|\Psi\right> =
\sum^{N_{\mathit{conf}}}_{\vec{n}}C_{\vec{n}}\left|\vec{n}\right>,
\eeq
where $|\vec{n}\rangle$ are permanents which are assembled as:
\beq\label{basic_permanents}
 \left|\vec{n}\right> =
\frac{1}{\sqrt{n_1!n_2!n_3!\cdots n_M!}} \left(b_1^\dag\right)^{n_1}\left(b_2^\dag\right)^{n_2}
\cdots\left(b_M^\dag\right)^{n_M}\left|vac\right>.
\eeq
$\vec{n}=(n_1,n_2,n_3,\cdots,n_M)$ represents the occupations of the orbitals
that preserve the total number of particles $n_1+n_2+n_3+\cdots+n_M=N$,
$\left|vac\right>$ is the vacuum
and $M$ is the number of the one-particle functions; for bosons $M \ge 1$.

The number of elements $N_{\mathit{conf}}$ in the bosonic state vector $\left|\Psi\right>$ is equal to the
number of the configurations used.
Here we consider the configurational space spanned by {\it all}
possible distributions of $N$ bosons over $M$ orbitals, i.e.,
a {\it complete} subspace of the respective configurational space.
We recall that one of the key consequences of this completeness is that an action of any operator on the state vector
results in a new state vector defined in the {\it same} configurational subspace.

The size of this complete Fock subspace is \cite{CC_paper}:
\beq\label{Nconf.b}
N_{\mathit{conf}}=\binom{N+M-1}{N}.
\eeq
This number is equal to the size of the configurational space spanned by $N$
spin-polarized fermions distributed over $M'=N+M-1$ fermionic orbitals, see section \ref{secII.A}.
Therefore, there exists one-to-one mapping between the configurational
spaces of the $N$-boson system distributed over $M$ bosonic orbitals
and a fermionic system made of $N$ fermions distributed over $M'=N+M-1$ fermionic orbitals and vise versa.
In other words, every fermionic configuration having $N+M-1$ components (occupation numbers) should be attributed
to a bosonic configuration characterized by $M$ components.
Let us compare these isomorphic bosonic and fermionic systems.
What is also equal in these two systems apart from the total number $N$ of particles?
The number of the fermionic unoccupied orbitals $M_v=M'-N$, i.e., the number of fermionic holes is
equal to the maximal number of the bosonic holes: $M_v=M-1$.
We note that the maximal number of bosonic holes appear in configurations like $\left| N,0,\cdots,0\right>$,
$\left| 0,N,0,\cdots,0\right>$, etc.
Since we already know how to enumerate fermionic configurations in terms of holes, see section \ref{secII.B},
we can adopt the fermionic enumeration scheme to the respective bosonic system.

\subsection{Mapping and enumeration of the permanents} \label{secIII.B}

Our goal is to provide simple and compact scheme for enumeration of the bosonic configurations.
To utilize the formal isomorphism between the bosonic and fermionic systems considered above
we first have to show how to map, or attribute a fermionic configuration to the bosonic one.
The rule is very simple:
the number of bosons residing in the first bosonic orbital $n_1$ is equal to the number of fermions
occupying successively the lowest fermionic orbitals from the bottom till the first fermionic hole.
In other words, the occupation
of the first bosonic orbital is defined as the number of ``1''s appearing in the $(N+M-1)$-component fermionic
vector [Eq.~(\ref{i-labeling})] up to the first hole, i.e., the first ``0'', when counting from the left.
The occupation number of the second bosonic orbital $n_2$
is defined as the number of ``1''s between the first and second ``0'', i.e., between the first and second fermionic holes.
The third bosonic occupation number $n_3$ is defined as the number of ``1''s between the second and third ``0'', and so on.
Table~\ref{table_fermions_vs_bosons} illustrates this mapping.

\begin{table}
\begin{tabular}{l|l|l}
Enumeration index $J$ & $N$ Fermions, $N+M-1$ orbitals & $N$ Bosons, $M$ orbitals \\
\hline
1 &$|\underbrace{1111\cdots111}_{N}\underbrace{000\cdots00}_{M-1}\rangle$& $|N,\underbrace{0,\cdots,0}_{M-1}\rangle$\\
2 &$|\underbrace{1111\cdots11}_{N-1}01\underbrace{000\cdots00}_{M-2}\rangle$& $|N-1,1,\underbrace{0,\cdots,0}_{M-2}\rangle$\\
$\cdots$&$\cdots$&$\cdots$\\
$J$ &$ |\underbrace{11111}_50\underbrace{111}_30\underbrace{1111}_40\cdots0\underbrace{111}_300\underbrace{111111}_6\rangle$&$|5,3,4,\cdots,3,0,6\rangle$\\
$\cdots$&$\cdots$&$\cdots$\\
$N_{\mathit{conf}}-1$&$|\underbrace{000\cdots00}_{M-2}10\underbrace{1111\cdots111}_{N-1}\rangle$& $|\underbrace{0,\cdots,0}_{M-2},1,N-1\rangle$\\
$N_{\mathit{conf}}$&$|\underbrace{000\cdots00}_{M-1}\underbrace{1111\cdots111}_N\rangle$& $|\underbrace{0,\cdots,0}_{M-1},N\rangle$\\
\end{tabular}
\caption{
Mapping between fermionic and bosonic configurations made of $N$ particles. The
number of the unoccupied fermionic orbitals is equal to the total number of the bosonic orbitals minus one: $M_v\equiv M-1$.
Now, we can adopt the fermionic enumeration scheme for bosons.
}
\label{table_fermions_vs_bosons}
\end{table}

Consequently, the bosonic occupation numbers and the positions of the fermionic holes are simply connected:
\begin{equation} \label{i-n-numbering}
\left. \begin{matrix}
n_1&=&i_1-1, \\
n_2&=&i_2-i_1-1,\\
\cdots \\       
n_k&=&i_k-i_{k-1}-1,\\
\cdots \\        
n_{M-1}&=&i_{M-1}-i_{M-2}-1,\\
n_{M}&=&N+M-i_{M-1}-1.\\
\end{matrix} \right.
\end{equation}
So, having a set of the bosonic occupation numbers $|n_1,n_2,n_3,\cdots,n_M \rangle$ using this scheme we can
restore the positions of the $M-1$ fermionic holes $(i_1,i_2,\cdots,i_{M-1})$ and the respective fermionic configuration and vise versa.
Now we can explicitly use the relations Eq.~(\ref{i-n-numbering}) between bosonic occupation numbers and fermionic holes
to uniquely address bosonic configurations utilizing Eq.~(\ref{i-numbering}).
Finally, omitting all intermediate steps we get the address of a generic bosonic configuration $|n_1,n_2,n_3,\cdots,n_M \rangle$
in the bosonic state vector:
\begin{equation} \label{j-numbering}
J(n_1,n_2,\cdots,n_{M-1},n_M)=1+\sum_{k=1}^{M-1} \binom{N+M-1-k-\sum_{l=1}^{k}n_l}{M-k}.
\end{equation}
Here $n_M$ enters the expression implicitly via the identity $N=n_1+n_2+n_3+\cdots+n_M$.

Let us summarize.
The system of $N$ bosons and $M$ bosonic orbitals spans the complete subspace of the Fock space of
$N_{\mathit{conf}}$ configurations [see Eq.~(\ref{Nconf.b})].
The dimension of any state vector $\left|\Psi\right>$ of the system is $N_{\mathit{conf}}$.
Every bosonic configuration in the respective Fock subspace is characterized by the set of $M$ occupation
numbers $|n_1,n_2,n_3,\cdots,n_M \rangle$.
We can attribute a unique address $J$ to each of the configurations according to the
rule Eq.~(\ref{j-numbering}): $J=J(n_1,n_2,\cdots,n_{M-1},n_M)$.
The proposed enumeration schemes Eq.~(\ref{i-numbering}) and Eq.~(\ref{j-numbering}) are equivalent
and connected via Eq.~(\ref{i-n-numbering}) to each other.
They can be equally applied to enumerate bosonic and fermionic configurations.
In other words, every many-body basis function written in the ``fermionic style'' -- holes position representation --
can be translated via Eq.~(\ref{i-n-numbering}) to the ``bosonic style''-- occupation numbers representation
$\left|n_1,n_2,\cdots,n_M\right>$.
However, the explicit use of the ``bosonic'' representation for fermions is neither appealing nor transparent.
Therefore, once Eq.~(\ref{j-numbering}) has been derived we preserve for the sake of clarity
the holes enumeration scheme for fermions
and occupation numbers enumeration scheme for bosons.

Now, we can represent the bosonic state vector $\left|\Psi\right>$ in a form similar to Eq.~(\ref{J_Psi.f}):
\beq\label{J_Psi.b}
\left|\Psi\right> =
\sum^{N_{\mathit{conf}}}_{J=1}C_{J}\left|J(n_1,n_2,\cdots,n_{M})\right>=\sum^{N_{\mathit{conf}}}_{J=1}C_{J}\left|J(\n)\right>,
\eeq
with the only difference that every bosonic configuration $|n_1,n_2,n_3,\cdots,n_M \rangle$ 
has a unique address $J$ which is computed according to Eq.~(\ref{j-numbering}), i.e., using bosonic occupation numbers
representation. The index $J$ runs over all $N_{\mathit{conf}}$ configurations.

\subsection{Applying operators to $\left|\Psi\right>$}\label{secIII.C}
Any bosonic operator can be represented as a sum of products of bosonic creation and annihilation operators.
If we now show that the action of any even combination of creation and destruction operators on a bosonic state vector Eq.~(\ref{J_Psi.b})
leads, as in the case of fermions, only to re-addressing (re-indexing) of the configurations (coordinates),
we can adopt all the ideas developed for fermions to bosonic systems.

Let us first consider a specific, say the $b^\dag_3b_2$ term which
kills a boson in the second bosonic orbital and creates a boson in the third one.
The result of the action of this pair of creation and annihilation operators on an arbitrary permanent
is known and reads:
$$
b^\dag_3b_2|n_1,n_2,n_3,\cdots,n_M\rangle=\sqrt{n_2}\sqrt{n_3+1}|n_1,n_2-1,n_3+1,\cdots,n_M\rangle.
$$
It acts on every many-body configuration having non-zero occupation $n_2$.
We can interpret this result as follows:
The operation of any even combination of bosonic creation and annihilation operators
on a permanent results in the re-addressing of this permanent to another one
multiplied with a trivial bosonic prefactor.
Clearly, all the ideas proposed for fermions in section \ref{secII} can be easily adopted here for bosons.

\subsubsection{The action of one- and two-body operators}\label{secIII.C.1}

To derive the results of the actions of bosonic one- and two-body operators on a state vector one can follow
the strategy used for fermions in subsection \ref{secII.C.1}.
We provide the final results and briefly discuss their meanings.
The action of one-body bosonic operator $b^\dag_kb_q$ on an incoming $\left|\Psi\right>$
results in $\left|\Psi^{kq}\right>$:
\beqn\label{one-body_oper.b}
\left|\Psi^{kq}\right>&\equiv& b^\dag_kb_q\left|\Psi\right> =
\sum^{N_{\mathit{conf}}}_{J=1}C^{kq}_{J}\left|J(\n)\right>, \\
C^{kq}_{J}&=&C_{J^{kq}}\sqrt{n_k}\sqrt{n_q+1}. \nonumber
\eeqn
Let us explain this expression.
According to Eq.~(\ref{j-numbering}) every configuration $\left|n_1,\cdots,n_k,\cdots,n_q,\dots,n_M \right>$
of the resulting vector $\left|\Psi^{kq}\right>$ has a unique address (index) $J =J(n_1,\cdots,n_k  ,\cdots,n_q  ,\dots,n_M)$.
The respective coefficient $C^{kq}_{J}$ is obtained as a product of
the $J^{kq}$-th component of the incoming vector $C_{J^{kq}}$
scaled by the bosonic $\sqrt{n_k}\sqrt{n_q+1}$ prefactor.
On the r.h.s. of Eq.~(\ref{one-body_oper.b}) we ``exchange''
the $k$-th and $q$-th indices due to a change of variables of the summation index $J$ applied,
similarly as it was done in Eq.~(\ref{one-body_oper.f}).
The index $J^{kq}$ is computed according to Eq.~(\ref{j-numbering})
and corresponds to the configuration $\left|n_1,\cdots,n_k-1,\cdots,n_q+1,\dots,n_M \right>$.
This simple and straightforward methodology is ideally suitable for programming.

The action of a general ($k\neq s \neq l \ne q$) two-body $b^\dag_kb^\dag_sb_lb_q$ term on
the incoming state vector $\left|\Psi\right>$ is defined as:
\beqn\label{two-body_oper.b}
\left|\Psi^{kslq}\right>&\equiv&b^\dag_kb^\dag_sb_lb_q\left|\Psi\right>=
\sum^{N_{\mathit{conf}}}_{J=1}C^{kslq}_{J}\left|J(\n)\right>, \\
C^{kslq}_{J}&=&C_{J^{kslq}}\sqrt{n_k} \sqrt{n_s}\sqrt{n_l+1} \sqrt{n_q+1}, \nonumber
\eeqn
where, after the change of variables of the summation index $J$,
the addresses of the incoming 
$J^{kslq}=J(n_1,\cdots,n_k-1,\cdots,n_s-1,\cdots,n_l+1,\cdots,n_q+1,\cdots,n_M)$
and resulting $J =J(n_1,\cdots,n_k,\cdots,n_s,\cdots,n_l,\cdots,n_q,\cdots,n_M)$ 
configurations are computed using Eq.~(\ref{j-numbering}).

\subsubsection{The action of the Hamiltonian}\label{secIII.C.2}

Using the same strategy as for the fermionic case we
group together the actions of all one-body and all two-body operators 
to get the total action of the Hamiltonian
on an initial bosonic state vector $\left|\Psi\right>$ [Eq.~(\ref{J_Psi.b})].

The one-body contributions read:
\beqn\label{H-psi_one-body}
\hat h \left|\Psi\right>& =&\sum_{k,q} h_{kq} \left [  b^\dag_kb_q \left|\Psi\right> \right ]
= \sum_{k,q} h_{kq} \left|\Psi^{kq}\right>=\sum^{N_{\mathit{conf}}}_{J=1}C^{\hat h}_{J}\left|J(\n)\right>,  \\
C^{\hat h}_J&=& \sum_{k,q} h_{kq}  C^{kq}_J. \nonumber
\eeqn
The two-body terms are:
\beqn\label{H-psi_two-body}
\hat W \left|\Psi\right>&=&\frac{1}{2}\sum_{k,s,q,l} W_{ksql} \left [b^\dag_k  b^\dag_s b_l b_q\left|\Psi\right> \right ]
=\frac{1}{2}\sum_{k,s,q,l} W_{ksql}  \left|\Psi^{kslq}\right>=\sum^{N_{\mathit{conf}}}_{J=1}C^{\hat W}_{J}\left|J(\n)\right>,  \\
C^{\hat W}_{J}&=&\frac{1}{2}\sum_{k,s,q,l} W_{ksql} C^{kslq}_J. \nonumber
\eeqn
Combining the contributions from the one- and two-body terms we get
the desired action of the Hamiltonian on a state vector $|\Psi\rangle$:
\beqn\label{Hpsi.b}
\hat H | \Psi \rangle&=&\hat h \left|\Psi\right>+\hat W \left|\Psi\right>=
\sum^{N_{\mathit{conf}}}_{J=1} C^{\hat H}_{J} \left|J(\n)\right>, \\
C^{\hat H}_{J}&=&C^{\hat h}_{J} + C^{\hat W}_{J}, \nonumber
\eeqn
which concludes our constructions for bosons.

\subsubsection{Other quantities of interest}\label{secIII.D}

We have seen above that the action of any bosonic operator on the state vector results in a new (resulting) state vector.
Similarly to fermionic systems, the expectation value of the respective operator
is immediately available as a dot-product of the incoming and resulting state vectors.
The matrix elements of the reduced one- and two- body density matrices as well as the expectation value of the Hamiltonian
can be obtained in a very similar way as have already been done for fermions in
Eqs.~(\ref{rho_ij.f},\ref{rho_ijkl.f}) and  Eq.~(\ref{psiHpsi}), respectively.
Other expectation values are also amenable to this formulation.

\section{The case of multi-component systems, binary mixtures}\label{secIV}

In this section we generalize the new ideas 
for effective and efficient operations in Fock space
proposed above for systems of spinless bosons and spin-polarized fermions
to multi-component systems. More specifically, we show that
the mapping and enumeration scheme discussed above
can be naturally extended to more general systems.

Let us considered binary mixtures with $N=N_A+N_B$ particles.
The mixture consists of $N_A$ identical particles (bosons or fermions)
of type $A$ and $N_B$ identical particles (bosons or fermions) of type $B$.
In what follows, we denote whenever needed the quantities in the mixture by $A$, $B$ or $AB$ superscripts.

The many-body Hamiltonian of the mixture has three kinds of terms:
\beqn\label{MB_hamil_mix}
 & &  \!\!\!\!\!\!\!\!\!\!\!\!\!\!\!\! \hat H^{(AB)} = \hat H^{(A)} + \hat H^{(B)} + \hat W^{(AB)}, \nonumber \\
 & &  \!\!\!\!\!\!\!\!\!\!\!\!\!\!\!\! \hat W^{(AB)} =
\sum_{k,k',q,q'} W^{(AB)}_{kk'qq'}  \hat a_k^\dag \hat a_q \hat b_{k'}^\dag \hat b_{q'} .
\eeqn
The first two terms of $\hat H^{(AB)}$ are the $A$ and $B$ single-species Hamiltonians
and can be read directly from Eq.~(\ref{MB_Ham}), formally replacing in the first case the $b_q$ and $b_k^\dag$ 
operators to $a_q$ and $a_k^\dag$, respectively.
The third term of $\hat H^{(AB)}$ is the interaction between the two species.
We call $\hat H^{(A)},\hat H^{(B)}$ intra- and $W^{(AB)}$ inter-species parts.

The many-body wavefunction $\Psi^{(AB)}$ is a linear combination of all possible products
of permutational-symmetry-adapted configurations:
\beq\label{MCTDH_XY_Psi}
\left|\Psi^{(AB)}\right> =
 \sum_{\vec{n},\vec{m}} C_{\vec{n}\vec{m}}(t)\left|\vec{n}\right> \times \left|\vec{m}\right>
\equiv \sum_{\vec{n},\vec{m}} C_{\vec{n}\vec{m}} \left|\vec{n},\vec{m}\right>.
\eeq
The configurations $\{\left|\vec{n}\right>\}$, $\{\left|\vec{m}\right>\}$
are either Slater determinants (\ref{basic_determinants}) or permanents (\ref{basic_permanents}),
depending on whether we deal with Bose-Bose, Fermi-Fermi, or Bose-Fermi mixtures.

In this work we make an assumption.
The total number of particles of each kind $N_A$ and $N_B$ is conserved.
In other words, there is no conversion between the particles, i.e., the particles of $A$ kind cannot become of $B$ kind and vise versa.
The Fock subspace of the $A$ subsystem is spanned by all possible permutations of $N_A$ particles
over $M_A$ orbitals ($N^A_{\mathit{conf}}$ configurations) and the configurational subspace of the $B$ subsystem
by permutations of $N_B$ particles over $M_B$ orbitals ($N^B_{\mathit{conf}}$ configurations).
Practically, in this case we work in the complete configurational subspaces,
i.e., the summation in Eq.~(\ref{MCTDH_XY_Psi}) over $\vec n$ runs from 1 till $N^A_{\mathit{conf}}$, and
over $\vec m$ from 1 till $N^B_{\mathit{conf}}$.
More strictly, the total configurational space is a tensor product of two complete subspaces
and any state vector of such a binary system has $N^A_{\mathit{conf}}N^B_{\mathit{conf}}$ components.

In the two previous sections \ref{secII} and \ref{secIII} we have seen that
enumeration schemes of the complete bosonic and fermionic configurational subspaces can be easily done by counting either holes Eq.~(\ref{i-numbering}) for fermions or particles Eq.~(\ref{j-numbering}) for bosons.
Therefore, if $J_A$ and $J_B$ label configurations in the $A$ and $B$ subspaces respectively, then the two-component
vector index $\vec{J}=\left( J^A,J^B \right)$ enumerates all possible configurations in the total product Fock subspace.
Now, we can represent the state vector of the mixture Eq.~(\ref{MCTDH_XY_Psi}) in a form where enumeration of the
configurations in the $A$ and $B$ subspaces are explicitly specified:
\beq\label{J_MCTDH_XY_Psi}
\left|\Psi^{(AB)}\right> =
 \sum^{N^A_{\mathit{conf}}}_{J_A=1} \sum^{N^B_{\mathit{conf}}}_{J_B=1}  C_{J_AJ_B}\left|J_A,J_B\right>
=\sum^{N^A_{\mathit{conf}}}_{J_A=1} \sum^{N^B_{\mathit{conf}}}_{J_B=1}  C_{J_AJ_B}\left|\vec J \right>.
\eeq
The key goal of this section is to demonstrate that the action of the total Hamiltonian 
Eq.~(\ref{MB_hamil_mix}) on the state vector of the mixture
can also be represented and computed without construction of the respective Hamiltonian matrix.

\subsection{General fermionic system}\label{secIV.A}

We consider the mixture of fermions of two different kinds $A$ and $B$.
Despite such an unusual abbreviature it is a standard system of spin-up and spin-down fermions where
the $A$ subsystem is associated with spin-up and $B$ with spin-down fermions.
Here, we consider the general, so called unrestricted case where
particles are not constrained to occupied the same spatial orbitals, i.e.,
the one-particle functions of the spin-up fermions can differ from the respective orbitals of the spin-down fermions. 
Such a treatment is allowed if the Hamiltonian does not have terms leading to
spin-flip phenomena, taking place for example in external magnetic fields.
The Hamiltonian Eq.~(\ref{MB_hamil_mix}) does not have such terms.

Let us first show that the action of typical one- and two-body terms contributing to
the $\hat H^{(A)}$,$\hat H^{(B)}$ as well as to the $\hat W^{(AB)}$ parts
on a basic configuration $\left|\vec J \right>\equiv \left|J_A(\i),J_B(\i')\right>$ 
translates or re-addresses it to another configuration,
multiplied by the respective fermionic prefactors.
Indeed, the configuration $\left|\vec J \right>$ 
is specified when the $M^A_v=M_A-N_A$ positions of the holes characterizing the $A$ subsystem, i.e., the $(i_1,i_2,\cdots,i_{M^A_v})$ set,
and the $M^B_v=M_B-N_B$ positions of the holes characterizing the $B$ subsystem, i.e., the $(i'_1,i'_2,\cdots,i'_{M^B_v})$ set, are given.
Then the $J_A(\i)$ and $J_B(\i')$ numbers available via Eq.~(\ref{i-numbering}) provide unique indices of the two-component address of this 
configuration in the state vector $\left|\Psi^{(AB)}\right>$.

If the initial holes of the $A$ subsystem are located at positions $(i_1,\cdots,k,\cdots,i_{M^A_v})$,
the action of the $\hat a_k^\dag \hat a_q$ operator kills a particle of $A$ kind in the $q$-th orbital and
creates the particle of $A$ kind in the $k$-th orbital -- we obtain the holes
at $(i_1,\cdots,q,\cdots,i_{M^A_v})$. Hence, according to Eq.~(\ref{i-numbering}) we get its new index $J^{kq}_A$
of the two-component address $(J^{kq}_A,J_B)$, where the standard fermionic prefactor has to be added
to account for the correct permutation symmetry: 
\beq\label{oper_A.mix.one-body}
\hat a_k^\dag \hat a_q  \left|J_A(\i),J_B(\i')\right>=(-1)^{d^{kq}_{J_A}} \left|J^{kq}_A,J_B\right>, k\in\i,q\not\in \i.
\eeq
Clearly, the actions of the two-body terms would also lead to re-addressings with some other fermionic prefactors:
\beqn\label{oper_A.mix.two-body}
\hat a_k^\dag \hat a_s^\dag  \hat a_l \hat a_q \left|J_A(\i),J_B(\i')\right>&=&
\hat a_s^\dag \hat a_l (-1)^{d^{kq}_{J_A}}  \left|J^{kq}_A,J_B\right> \\
&=& (-1)^{d^{kq}_{J_A}} (-1)^{d^{sl}_{J^{kq}_A}} \left|J^{kslq}_A,J_B\right>,  k,s\in\i, \,\, l,q\not\in \i. \nonumber
\eeqn
The action of creation and annihilation operators from the $A$ subspace changes the address of $J_A$ component only and 
does not touch the $J_B$ ones.
Similarly, the actions of the one- and two-body $B$ terms of the Hamiltonian read:
\beqn\label{oper_B.mix}
\hat b_{k'}^\dag \hat b_{q'} \left|J_A(\i),J_B(\i')\right>&=&(-1)^{d^{k'q'}_{J_B}} \left|J_A,J^{k'q'}_B\right>, k'\in\i',q'\not\in\i',\\
\hat b_{k'}^\dag \hat b_{s'}^\dag  \hat b_{l'} \hat b_{q'} \left|J_A(\i),J_B(\i')\right>&=&
\hat b_{s'}^\dag \hat b_{l'}  (-1)^{d^{k'q'}_{J_B}}  \left|J_A,J^{k'q'}_B\right> \\
&=&
(-1)^{d^{k'q'}_{J_B}} (-1)^{d^{s'l'}_{J^{k'q'}_B}} \left|J_A,J^{k's'l'q'}_B\right>, k',s'\in\i', \,\, l',q'\not\in \i'. \nonumber
\eeqn
Now, we show that the action of the inter-species terms from the $ \hat W^{(AB)}$ part also
results in a re-addressing of the initial configuration to another one with a different fermionic prefactor:
\beqn\label{oper_AB.mix}
\hat a_k^\dag \hat a_q  \hat b_{k'}^\dag \hat b_{q'}  \left|J_A(\i),J_B(\i')\right> 
=(-1)^{d^{kq}_{J_A}} (-1)^{d^{k'q'}_{J_B}} \left|J^{kq}_A,J^{k'q'}_B\right>, k\in\i, q\not\in \i, k'\in\i',q'\not\in\i'.
\eeqn

We have shown that the action of each term of the Hamiltonian Eq.~(\ref{MB_hamil_mix}) on a general configuration
re-addresses it to another one within the same Fock subspace.
This allows us to conclude that the action of the total Hamiltonian Eq.~(\ref{MB_hamil_mix})
on a state vector Eq.~(\ref{J_MCTDH_XY_Psi}) can be obtained directly without construction of the respective Hamiltonian matrix:
\beqn\label{Hpsi.AB.ff}
\hat H^{(AB)} \left|\Psi^{(AB)}\right>&=&
\sum^{N^A_{\mathit{conf}}}_{J_A=1} \sum^{N^B_{\mathit{conf}}}_{J_B=1} C^{\hat H^{(AB)}}_{J_AJ_B} \left|J_A(\i),J_B(\i')\right>,\\
C^{\hat H^{(AB)}}_{J_AJ_B}&=&
C^{\hat H^{(A)}}_{J_AJ_B}+
C^{\hat H^{(B)}}_{J_AJ_B}+
C^{\hat W^{(AB)}}_{J_AJ_B}, \nonumber
\eeqn
where $C^{\hat H^{(A)}}$ and $C^{\hat H^{(A)}}$ can be read from Eqs.~(\ref{one-body_oper.f},\ref{two-body_oper.f},\ref{H-psi_one-body.f},\ref{H-psi_two-body.f},\ref{Hpsi.f}),
and the $C^{\hat W^{(AB)}}_{J_AJ_B}$ can be easily derived using Eq.~(\ref{oper_AB.mix}):
\beqn\label{two-body_mix.mix}
\hat W^{(AB)} \left|\Psi^{(AB)}\right>&=&
\sum_{k,k',q,q'} \hat W^{(AB)}_{kk'qq'} \hat a_k^\dag \hat a_q \hat b_{k'}^\dag \hat b_{q'} \left|\Psi^{(AB)}\right>
=\sum^{N^A_{\mathit{conf}}}_{J_A=1} \sum^{N^B_{\mathit{conf}}}_{J_B=1}
C^{\hat W^{(AB)}}_{J_AJ_B} \left|J_A(\i),J_B(\i')\right>, \\
C^{\hat W^{(AB)}}_{J_AJ_B}& =& \sum_{k,k',q,q'}\hat W^{(AB)}_{kk'qq'} C^{kk'qq'}_{J_AJ_B}, \nonumber \\ 
C^{kk'qq'}_{J_AJ_B} &=&
\left \{ \begin{matrix}
C_{J^{kq}_{A} J^{k'q'}_{B}} (-1)^{d^{kq}_{J_A}} (-1)^{d^{k'q'}_{J_B}} ; \,k\not\in\i, q\in\i,  k'\not\in\i', q'\in \i'\\
0; \,\, {\mathit{otherwise}} .\nonumber
\end{matrix} \right.
\eeqn
Each element of the resulting vector has a unique address $(J_A(\i),J_B(\i'))$ 
characterized by the two sets of holes $(i_1,\cdots,q,\cdots,i_{M^A_v})$ and
$(i'_1,\cdots,q',\cdots,i'_{M^B_v})$, i.e., $J_A$ and $J_B$ are obtained by using Eq.~(\ref{i-numbering}).
The value of the element $C^{\hat W^{(AB)}}_{J_AJ_B}$
is computed as a sum of the $C_{J^{kq}_{A} J^{k'q'}_{B}}$ components of the incoming state vector scaled by the respective integrals $W^{(AB)}_{kk'qq'}$ and fermionic prefactors.
The address $(J^{kq}_{A}$, $J^{k'q'}_{B})$ of each of these components is obtained for every given set $k,k,'q,q'$
by applying Eq.~(\ref{i-numbering}) to $(i_1,\cdots,k,\cdots,i_{M^A_v})$ and $(i'_1,\cdots,k',\cdots,i'_{M^B_v})$.
Finally, the expectation value of the Hamiltonian as well as of any other operator
are available as a dot-product of the incoming and respective resulting vectors in 
a very similar manner as it has been done in the single-component case, Eq.~(\ref{psiHpsi}).

\subsection{Mixture of bosons}\label{secIV.B}

Here, we consider the mixture of bosons of two different kinds, $A$ and $B$. 
In section \ref{secIII} we have seen that operations with single-component bosonic systems
can be done without representing the respective operators in the matrix form.
Now, we demonstrate the usefulness of this theory and ideas to bosonic mixtures.

The general configuration $\left|\vec J \right>\equiv \left|J_A(\n),J_B(\n')\right>$ is specified by the
bosonic occupation numbers corresponding to the first $(n_1,n_2,\cdots,n_{M_A})$ 
and second $(n'_1,n'_2,\cdots,n'_{M_B})$ bosonic subsystems.
Let us first show that the actions of typical intra-species terms from the $A$ subsystem lead to a re-addressings:
\beq
\label{oper_A.mix.bb.one-body}
\hat a_k^\dag \hat a_q  \left|J_A(\n),J_B(\n')\right>=\sqrt{n_k+1}\sqrt{n_q} \left|J^{kq}_A,J_B\right>,
\eeq
\beqn
\label{oper_A.mix.bb.two-body}
\hat a_k^\dag \hat a_s^\dag  \hat a_l \hat a_q \left|J_A(\n),J_B(\n')\right>
&=&\hat a_s^\dag \hat a_l \sqrt{n_k+1}\sqrt{n_q} \left|J^{kq}_A,J_B\right> \\
&=& \sqrt{n_k+1}\sqrt{n_q} \sqrt{n_s+1}\sqrt{n_l} \left|J^{kslq}_A,J_B\right> \nonumber .
\eeqn
Similar expressions can be obtained for the action of the intra-species terms associated with the $B$ subsystem.
Clearly, the inter-bosonic term acting on a general configuration translates it to another
one, weighted by the respective bosonic prefactor:
\beqn\label{oper_AB.mix.bb}
\hat a_k^\dag \hat a_q  \hat b_{k'}^\dag \hat b_{q'}  \left|J_A(\n),J_B(\n')\right> 
= \sqrt{n_k+1}\sqrt{n_q}  \sqrt{n_{k'}+1}\sqrt{n_{q'}}  \left|J^{kq}_A,J^{k'q'}_B\right>.
\eeqn

Having verified that the actions of intra- and inter-species terms of the bosonic operators on a general configuration
result in its re-addressing with some scaling prefactors, we can explicitly express the result of the action of the Hamiltonian
on a state vector Eq.~(\ref{J_MCTDH_XY_Psi}), i.e., to sum up the contributions from all the terms of Eq.~(\ref{MB_hamil_mix}):
\beqn\label{Hpsi.AB.bb}
\hat H^{(AB)} \left|\Psi^{(AB)}\right>&=&
\sum^{N^A_{\mathit{conf}}}_{J_A=1} \sum^{N^B_{\mathit{conf}}}_{J_B=1} 
C^{\hat H^{(AB)}}_{J_AJ_B} \left|J_A(\n),J_B(\n')\right> \\
C^{\hat H^{(AB)}}_{J_AJ_B}&=&
C^{\hat H^{(A)}}_{J_AJ_B}+
C^{\hat H^{(B)}}_{J_AJ_B}+
C^{\hat W^{(AB)}}_{J_AJ_B}, \nonumber
\eeqn
where $C^{\hat H^{(A)}}$ and $C^{\hat H^{(A)}}$ can be deduced from Eqs.~(\ref{one-body_oper.b},\ref{two-body_oper.b},\ref{H-psi_one-body},\ref{H-psi_two-body},\ref{Hpsi.b}),
and the $C^{\hat W^{(AB)}}_{J_AJ_B}$ can be derived using Eq.~(\ref{oper_AB.mix.bb}) in a very similar way
as done in Eq.~(\ref{two-body_mix.mix}).

We conclude that the operation of the Hamiltonian representing a binary mixture of bosons on a state vector
can be performed directly without constructing the respective Hamiltonian matrix.

\subsection{Mixture of bosons and fermions}\label{secIV.C}

In the two preceding subsections we have seen that effective and efficient operations and manipulations with
state vectors of a binary mixture of bosons or a binary mixture of fermions are possible without resorting to the matrix
representation of the respective operators.
Actually, the final equations ~(\ref{Hpsi.AB.ff}) and (\ref{Hpsi.AB.bb}) for $\hat H^{(AB)} \left|\Psi^{(AB)}\right>$
are almost identical, with the only exception of the enumeration scheme used to address fermionic and bosonic configurations.
In the Fermi-Fermi case we use the positions of the holes $\i,\i'$ of the $A$ and $B$ subsystems
to specify the fermionic configurations and Eq.~(\ref{i-numbering}) to compute the indices of the two-component address
of the mixture while in the Bose-Bose case we utilize bosonic occupation numbers $\n,\n'$ to
specify the bosonic configurations and Eq.~(\ref{j-numbering}) to compute the components of the respective address.
The goal of this subsection is to demonstrate, for completeness, the validity and applicability
of the theory to a mixed system of bosons and fermions.

Let $A$ be the fermionic subsystem and $B$ -- the bosonic one.
Now, to specify the general configuration  $\left|\vec J \right>\equiv \left|J_A(\i),J_B(\n)\right>$ 
one has to provide a set of the fermionic holes $(i_1,i_2,\cdots,i_{M^A_v})$
and a set of the bosonic occupation numbers $(n_1,n_2,\cdots,n_{M_B})$. The two-component address of this configuration 
is defined by the two numbers $J_A(\i)$ and $J_B(\n)$ computed by using Eq.~(\ref{i-numbering}) and Eq.~(\ref{j-numbering}), respectively.

Clearly, the operation of the one- and two-body fermionic terms on the general configuration 
changes the positions of the fermionic holes and does not affect the bosonic occupation numbers.
In other words, only the first (fermionic) index of the two-component address $\left|\vec J \right>$
is changed, like in Eqs.~(\ref{oper_A.mix.one-body},\ref{oper_A.mix.two-body}). Similarly, the action of the pure bosonic terms
on a general configuration re-addresses only the bosonic index, analogously to Eqs.~(\ref{oper_A.mix.bb.one-body},\ref{oper_A.mix.bb.two-body}).
The Bose-Fermi interaction terms lead to the change of both fermionic and bosonic parts of the two-component address:
\beqn\label{oper_AB.mix.fb}
\hat a_k^\dag \hat a_q  \hat b_{k'}^\dag \hat b_{q'}  \left|J_A(\i),J_B(\n)\right> 
= (-1)^{d^{kq}_{J_A}}   \sqrt{n_{k'}+1}\sqrt{n_{q'}}  \left|J^{kq}_A,J^{k'q'}_B\right>, k\in\i, q\not\in \i.
\eeqn
Every configuration characterized by $\i=(i_1,i_2,\cdots,i_{M^A_v})$ and $\n=(n_1,n_2,\cdots,n_{M_B})$
and having, therefore, the address $(J_A(\i),J_B(\n))$ is translated to a configuration $(J^{kq}_A,J^{k'q'}_B)$
where the $k$-th fermionic hole was filled and a new hole at $q$ has been created, i.e., $(i_1,\cdots,q,\cdots,i_{M^A_v})$,
and simultaneously one boson from bosonic orbital $q'$ has been transfered to orbital $k'$, i.e., $(n_1,\cdots,n_{k'}+1,\cdots,n_{q'}-1,\cdots,n_{M_B})$.

The action of any term of the Bose-Fermi Hamiltonian on a general configuration leads to its translation, i.e., re-addressing
with some known prefactor. This allows us to apply the  developed theory and write down the result of the action of the
Hamiltonian on a state vector of the system as follows:
\beqn\label{Hpsi.AB.bf}
\hat H^{(AB)} \left|\Psi^{(AB)}\right>&=&
\sum^{N^A_{\mathit{conf}}}_{J_A=1} \sum^{N^B_{\mathit{conf}}}_{J_B=1}
C^{\hat H^{(AB)}}_{J_AJ_B} \left|J_A(\i),J_B(\n)\right> ,\\
C^{\hat H^{(AB)}}_{J_AJ_B}&=&
C^{\hat H^{(A)}}_{J_AJ_B}+
C^{\hat H^{(B)}}_{J_AJ_B}+
C^{\hat W^{(AB)}}_{J_AJ_B} \nonumber,
\eeqn
where the contributions $\hat C^{H^{(A)}}$ and $\hat C^{H^{(B)}}$ from the actions of the intra-species Hamiltonians
can be deduced from
Eqs.~(\ref{one-body_oper.f},\ref{two-body_oper.f},\ref{H-psi_one-body.f},\ref{H-psi_two-body.f},\ref{Hpsi.f}) and
Eqs.~(\ref{one-body_oper.b},\ref{two-body_oper.b},\ref{H-psi_one-body},\ref{H-psi_two-body},\ref{Hpsi.b}), respectively.
The results of the action of the Bose-Fermi interactions $C^{W^{(AB)}}_{J_AJ_B}$
can be derived using Eq.~(\ref{oper_AB.mix.fb}) in a very similar way
as done in Eq.~(\ref{two-body_mix.mix}).

\subsection{More components}\label{secIV.D}

In the previous subsections we have shown that the total configurational spaces of quantum systems made of two kinds of particles, i.e.,
Fermi-Fermi, Bose-Bose and Bose-Fermi mixtures can be labeled by two-component vector index $\vec{J}=\left( J_A,J_B \right)$.
A system with a larger number of components can be addressed by a multi-component vector index
$\vec{J}=\left( J_A,J_B,J_C,\cdots\right)$.
Then, all the experience collected in this paper can be expanded to these systems as well.
The only constrain is that the total number of particles of each kind is conserved, i.e.,
there are no terms in the Hamiltonian leading to particle conversion.
We recall that, depending on the quantum statistics of the subsystem, i.e., whether we are dealing with fermions or bosons,
we apply different schemes to enumerate the respective configurations.
Both derived enumeration schemes, i.e., Eq.~(\ref{i-numbering}) for fermions and Eq.~(\ref{j-numbering}) for bosons
are applicable only if $N_A=const,N_B=const,N_C=const,\ldots$.
We just mention here that a generalization of the presented enumeration
schemes can be adopted also to systems with particle conversion.
This issue is out of the scope of the present study.

The most relevant conclusion is that for multi-component Hamiltonians the action of each term on a state vector
leads to re-addressing of the configurations with some quantum-statistics-dependent, but simple and known prefactors.
We can find the results of actions of each of these terms on the state vector independently.
The total action of the Hamiltonian is obtained by summing over all the resulting vectors.
Having at hand the initial and resulting state vectors we can readily compute
the expectation value of the Hamiltonian as a simple dot-product of these two vectors.

\section{Remarks on implementation}\label{secV}

Here, to discuss a strategy for a practical implementation of the ideas presented in all previous sections
we refer to the single-component bosonic or fermionic system.
To define the system means to specify the number of particles $N$, their quantum statistics, and the number of the orbitals $M$.
The integrals $h_{kq}$ and $W_{ksql}$ specify the Hamiltonian or any other operator of interest.
Depending on the type of the quantum statistics the $N$ and $M$ define the length of the state vector $N_{\mathit{conf}}$ according
to Eq.~(\ref{Nconf.f}) or Eq.~(\ref{Nconf.b}), respectively.
It is equal to the number of the elements in the one-dimensional vector-array of complex numbers 
$\{C_{J}\}^{N_{\mathit{conf}}}_{J=1}$ representing
the state vector $\left|\Psi\right>$ of the system.

Now we explain basic computational steps needed to get the result of the action of the Hamiltonian on a state vector.
As a first step, for a given incoming state vector $\left|\Psi\right>$, i.e., a vector-array of expansion coefficients $\{C_{J}\}^{N_{\mathit{conf}}}_{J=1}$,
the action of every pair or quartet of creation and annihilation operators
is evaluated using Eqs.~(\ref{one-body_oper.f},\ref{two-body_oper.f}) for fermionic
or Eqs.~(\ref{one-body_oper.b},\ref{two-body_oper.b}) for bosonic systems.
The result of the action of each such an operator on a state vector is another state vector, i.e.,
as an outcome we get again a one-dimensional vector-array of some other complex numbers
$\{C'_{J}\}^{N_{\mathit{conf}}}_{J=1}$ of the same length $N_{\mathit{conf}}$.
It is very important to stress that the action of each of these operators can be computed independently,
implying an effective parallelization.
For example, each available computational node is designated to a specific pair or quartet of creation and annihilation operators.
Next, the incoming array $\{C_{J}\}^{N_{\mathit{conf}}}_{J=1}$
is broadcasted to every node and the respective actions take place producing the resulting arrays.
We apply Eqs.~(\ref{rho_ij.f},\ref{rho_ijkl.f}) on each node to compute the corresponding element
of the reduced one- or two-body density matrices as dot-products of the incoming and respective resulting vector-arrays.
Then, by multiplying each and every elements of the resulting vector-array by the corresponding integral $h_{kq}$ or $W_{ksql}$
we get on each node the desired action of the respective Hamiltonian's term.
Now to sum up the resulting vectors from all the nodes we can use an appropriate collective operation
and get the desired total action of the Hamiltonian on the incoming state vector.
Finally, using Eq.~(\ref{psiHpsi}) we easily compute the expectation value of the
Hamiltonian $\langle \Psi| \hat H | \Psi \rangle$ as a dot-product
of the total resulting and incoming state vectors.

This technique has several advantages. First of all, it does not require the evaluation of the Hamiltonian matrix elements in the given
many-body basis set. Consequently, there is no need to construct, store and address these elements of the Hamiltonian matrix at all.
Second, the elements of the reduced one and two-body matrices are immediately and naturally available.
Third, this technique can be easily extended to three- and higher-body interactions potentials.
Last but not the least, such a strategy implies very effective parallelization schemes, which are of high demands
in modern computational physics.
Finally, in the formulation of this method we did not specify explicitly the $h_{kq}$ and $W_{ksql}$ numbers, therefore, it is valid
for general many-body Hamiltonians or for any other operators represented in the second quantization form.
Consequently, this scheme can be successfully applied to standard real-space Hamiltonians as well as to discrete ones,
e.g., of the Bose-Hubbard type.
The derivations of the re-addressing scheme has been done for the complete Fock subspace spanned
by permutation of $N$ particles over $M$ orbitals, but in principle, any selected subspace can be used.
For example, an implementation of additional constraints on possible excitation patterns for fermions
or restrictions on occupancies of the higher bosonic orbitals
lead to considerable reduction of the respective configurational subspaces.
In these cases the enumeration schemes derived above have to be modified accordingly.
Moreover, one has to pay additional attention to the re-addressing cases leading beyond the selected configurational subspace.
This opens, on the other hand, a new vision on size-consistency issues \cite{CI,CC_paper} -- it can be now explicitly considered and analyzed.

\section{Summary and conclusions}\label{secVI}
In this paper we provide a novel, effective and general technique to construct the result of the action of any operator 
represented in the second quantized form on a many-body state vector.
Within a standard framework one represents the corresponding operator in a matrix form
and obtains the desired action of the operator by applying a matrix-to-vector multiplication.
We have shown that the same result can be reached without even referring to the respective matrix elements.
Considering configurations as coordinates of the many-body state vector,
we first demonstrated that the action of any even combination of creation and annihilation operators on
a configuration translates or re-addresses it to another configuration.
In other words, we have seen that such an action is equivalent to permutation of coordinates of the initial state vector,
weighted by some trivial prefactors.
The total action of any operator, represented in the second quantized form on a state vector is a sum of the actions of all its terms.

Next, for a complete subspace of the configurational space spanned by permutation of $N$ fermions over $M$ orbitals we present 
a simple and compact scheme to enumerate the fermionic configurations according to the given set of the hole's positions.
Then, using the formal isomorphism between the configurational spaces formed by this fermionic system 
and the system of $N$ bosons distributed over the corresponding number of bosonic orbitals
we invent a simple and compact scheme to enumerate bosonic configurations according to the given set of the bosonic occupation numbers.
Using these enumeration algorithms we directly construct the result of the
action of any pair or quartet of the annihilation and creation operators
on a state vector. Moreover, the respective matrix elements of the reduced one- and two-body density matrices are naturally available 
as simple dot-products of the initial and resulting state vectors.
This allows us to combine the total action of the Hamiltonian as a sum of all its terms into simple and compact formulas,
which can be directly implemented and permit straightforward parallelization.
The proposed ideas to operate with fermionic and bosonic Hamiltonians have been directly extended to binary mixtures
of fermions and of bosons as well as to the Bose-Fermi mixtures.
We have also shown that the same ideas can be naturally applied to systems made of a larger number of components as well.

Finally, since the $\hat H \left|\Psi\right>$ is the basic building block appearing in the computations of statical properties as well as
of the evolution dynamics of many-body systems, we expect that the implementation of the theory into
the respective computational approaches will increase their efficiency and enable applications to systems
made of larger number of particles than currently possible.
Indeed, the proposed ideas have recently been implemented for bosonic systems 
\cite{BJJ_arXiv,to_be_submitted} within the multiconfigurational 
time-dependent Hartree for bosons (MCTDHB) \cite{MCTDHB}, and 
applications to multi-boson long-time dynamics in double-well 
\cite{BJJ_arXiv} and triple-well \cite{to_be_submitted} traps have 
already been performed successfully.

\begin{acknowledgments}
Financial support by the Deutsche Forschungsgemeinschaft is gratefully acknowledged.
\end{acknowledgments}

\end{document}